\magnification1200


\vskip 2cm
\centerline
{\bf   A sketch of brane dynamics in seven and eight dimension using E theory}
\vskip 1cm
\centerline{ Peter West}
\centerline{Department of Mathematics}
\centerline{King's College, London WC2R 2LS, UK}
\vskip 2cm
\leftline{\sl Abstract}
Using the general properties that have emerged from  E theory we sketch the generic features of  the dynamics of branes in seven and eight dimensions. The dynamical equations are a set of duality equations involving  the coordinates of the vector representation of $E_{11}$.

\vskip2cm
\noindent

\vskip .5cm

\vfill
\eject
{\bf 1 Introduction}
\medskip
The dynamics in E theory follows from  the non-linear realisation of $E_{11}\otimes _s l_1$. The $E_{11}$ part encodes the fields and the vector ($l_1$) representation encodes  the coordinates. The non-linear realisation is constructed from a group element $g$ which belongs to the group 
${\cal E}_l$, whose Lie algebra is $E_{11}\otimes _s l_1$,  and it is subject to the transformations $g\to g_0 g$ where the rigid transformation $g_0\in {\cal E}_l$ and $g\to gh$ where the local transformation $h$ is in the local subgroup which is specified as part of the definition of the non-linear realisation. The dynamics is determined by requiring that it is invariant under these two transformations and so different choices of local subalgebra lead to different dynamics. The symmetries in the local subgroup correspond to symmetries that are preserved and so linearly realised, while   the ones not in the local subalgebra are those that are spontaneously broken and these are non-linearly realised. 
\par
If we take the fields to depend on the coordinates and the local subalgebra to be the Cartan involution invariant subalgebra of $E_{11}$, denoted $I_c(E_{11})$ then one derives the low energy effective action for strings and branes as conjectured long ago [1,2]. Indeed, once one restricts to the lowest level fields and coordinates, the dynamics contains the maximal supergravity theories. In particular if one takes the decomposition of $E_{11}$ to GL(11) one finds the equations of motion of eleven dimensional supergravity [3,4] and one will inevitably find the other maximal supergravities if one takes the  other decompositions corresponding to the algebra that results from deleting the other nodes in the $E_{11}$ Dynkin diagram. For a review see reference [5].  

However, if one takes the coordinates to depend on variables that parameterise the branes,  the fields to depend on these coordinates and the local subgroup ${\cal H}$ to be a subalgebra of  $I_c(E_{11})$  then one finds the dynamics of branes. The different choices of local subgroup leads to the different branes. However, each brane carries the full $E_{11}$ symmetry but  the symmetries  that are spontaneously broken vary  from brane to brane [6,7]. 
\par
The decomposition of $E_{11}$   to GL(11) results in the theory in eleven dimensions and the generators of $E_{11}$ can  be found, for example,  in the book [8], while  the generators in the vector representation are given by [2,9,10 ]
$$
P_{\underline a}, Z^{\underline a_1 \underline a_2}, \ Z^{\underline a_1\ldots \underline a_5}, \ Z^{\underline a_1\ldots \underline a_7,\underline b},\  Z^{\underline a_1\ldots \underline a_8},\ 
\  Z^{ \underline b_1 \underline b_2  \underline b_3,  \underline a_1\ldots  \underline a_8},\ 
\  Z^{( \underline c  \underline d ),  \underline a_1\ldots  \underline a_9},\ 
\  Z^{ \underline c  \underline d, \underline a_1\ldots  \underline a_9},\ 
$$
$$
\  Z^{ \underline c, \underline a_1\ldots  \underline a_{10}}\ (2),\ 
Z^{\underline a_1\ldots \underline a_{11}} ,\ 
Z^{ \underline c,   \underline d_1\ldots   \underline d_4, \underline a_1\ldots  \underline a_9},\ 
Z^{ \underline c_1\ldots  \underline c_6, \underline a_1\ldots  \underline a_8},\ 
Z^{ \underline c_1\ldots  \underline c_5, \underline a_1\ldots  \underline a_9},\ 
Z^{ d_1, \underline c_1  \underline c_2  \underline c_3, \underline a_1\ldots  \underline a_{10}},\ (2),
$$
$$
\  Z^{ \underline c_1 \ldots  \underline c_4, \underline a_1\ldots  \underline a_{10}},\ (2),\  Z^{( \underline c_1 \underline c_2 \underline c_3 ),\underline a_1\ldots \underline a_{11}},\ Z^{ \underline c, \underline b_1 \underline b_2, \underline a_1\ldots \underline a_{11}},\ (2),\  
Z^{ \underline c_1\ldots  \underline c_{3},\underline a_1\ldots \underline a_{11}},\ (3),\ 
 \ldots  
\eqno(1.1)$$
\par
Each block of  indices contain indices that are totally antisymmetrised except when  $()$ is present and this  indicates that the indices are symmetrised instead. The elements have  multiplicity one except when there is a bracket after the object which contains a number that gives the multiplicity. All the generator belong to irreducible representations of SL(11), for example $ Z^{ \underline b_1 \underline b_2  \underline b_3,  \underline a_1\ldots  \underline a_8}$ obeys the constraint $Z^{ \underline b_1 \underline b_2  [\underline b_3,  \underline a_1\ldots  \underline a_8]}=0$. 
\par
The theory in $D$ dimensions results  from deleting the node $D$ in the $E_{11}$ Dynkin diagram and the generators in the vector representation which have totally antisymmetrised indices are given in the table below [11, 6,10]
\medskip
\eject
{\centerline{\bf {Table 1. The form generators  in the $l_1$ representation  in D
dimensions}}}
\medskip
$$\halign{\centerline{#} \cr
\vbox{\offinterlineskip
\halign{\strut \vrule \quad \hfil # \hfil\quad &\vrule Ê\quad \hfil #
\hfil\quad &\vrule \hfil # \hfil
&\vrule \hfil # \hfil Ê&\vrule \hfil # \hfil &\vrule \hfil # \hfil &
\vrule \hfil # \hfil &\vrule \hfil # \hfil &\vrule \hfil # \hfil &
\vrule \hfil # \hfil &\vrule#
\cr
\noalign{\hrule}
D&G&$Z$&$Z^{{\underline a}}$&$Z^{{\underline a}_1{\underline a}_2}$&$Z^{{\underline a}_1\ldots {\underline a}_{3}}$&$Z^{{\underline a}_1\ldots {\underline a}_
{4}}$&$Z^{{\underline a}_1\ldots {\underline a}_{5}}$&$Z^{{\underline a}_1\ldots {\underline a}_6}$&$Z^{{\underline a}_1\ldots {\underline a}_7}$&\cr
\noalign{\hrule}
8&$SL(3)\otimes SL(2)$&$\bf (3,2)$&$\bf (\bar 3,1)$&$\bf (1,2)$&$\bf
(3,1)$&$\bf (\bar 3,2)$&$\bf (1,3)$&$\bf (3,2)$&$\bf (6,1)$&\cr
&&&&&&&$\bf (8,1)$&$\bf (\bar 6,2)$&$\bf (18,1)$&\cr Ê&&&&&&&$\bf (1,1)$&&$
\bf
(3,1)$&\cr Ê&&&&&&&&&$\bf (6,1)$&\cr
&&&&&&&&&$\bf (3,3)$&\cr
\noalign{\hrule}
7&$SL(5)$&$\bf 10$&$\bf\bar 5$&$\bf 5$&$\bf \overline {10}$&$\bf 24$&$\bf
40$&$\bf 70$&-&\cr Ê&&&&&&$\bf 1$&$\bf 15$&$\bf 50$&-&\cr
&&&&&&&$\bf 10$&$\bf 45$&-&\cr
&&&&&&&&$\bf 5$&-&\cr
\noalign{\hrule}
6&$SO(5,5)$&$\bf \overline {16}$&$\bf 10$&$\bf 16$&$\bf 45$&$\bf \overline
{144}$&$\bf 320$&-&-&\cr &&&&&$\bf 1$&$\bf 16$&$\bf 126$&-&-&\cr
&&&&&&&$\bf 120$&-&-&\cr
\noalign{\hrule}
5&$E_6$&$\bf\overline { 27}$&$\bf 27$&$\bf 78$&$\bf \overline {351}$&$\bf
1728$&-&-&-&\cr Ê&&&&$\bf 1$&$\bf \overline {27}$&$\bf 351$&-&-&-&\cr
&&&&&&$\bf 27$&-&-&-&\cr
\noalign{\hrule}
4&$E_7$&$\bf 56$&$\bf 133$&$\bf 912$&$\bf 8645$&-&-&-&-&\cr
&&&$\bf 1$&$\bf 56$&$\bf 1539$&-&-&-&-&\cr
&&&&&$\bf 133$&-&-&-&-&\cr
&&&&&$\bf 1$&-&-&-&-&\cr
\noalign{\hrule}
3&$E_8$&$\bf 248$&$\bf 3875$&$\bf 147250$&-&-&-&-&-&\cr
&&$\bf1$&$\bf248$&$\bf 30380$&-&-&-&-&-&\cr
&&&$\bf 1$&$\bf 3875$&-&-&-&-&-&\cr
&&&&$\bf 248$&-&-&-&-&-&\cr
&&&&$\bf 1$&-&-&-&-&-&\cr
\noalign{\hrule}
}}\cr}$$
\medskip
 We note the presence of the generators at level one that are Lorentz scalars but  belong to representations of  $E_{11-D}$. Given the one to one relation between generators in the vector representation and coordinates in the non-linear realisation it follows that  for each element in the table we have  a  coordinates in the spacetime through which the brane moves. As such the  coordinates in the above table are  the main characters in this paper. In fact  there are an infinite number of coordinates in the vector representation most of which have indices that can not be written as a single antisymmetrised block. These later coordinates will not play a significant role in this paper as we will be concerned with low level branes.  The coordinates in the vector representation at higher levels than the usual coordinates $x^a$ of our familiar spacetime play an essential role in the construction of the low energy effective action of strings and branes even though they must be truncated out to gain the supergravity results we are familiar with. The Lorentz scalar coordinates, in the first column first proposed in the  $E_{11}$ papers referenced above,  are the starting point for   papers on the so called exceptional field theory, see for example reference [12] for an account of this.  As we will discover, while some coordinates given the embedding of the brane in our usual spacetime, some of the coordinates are the world volume fields of the branes [6,7]. 
\par
We now briefly review some of the main features of how one computes the brane dynamics from the Cartan forms [6,7]. While this paper does not contain any detailed calculations the construction  below is required to justify  the discussions in this paper. 
We can write the group element $g$ of the  non-linear realisation  in the form
$
g=g_l g_h g_E $ where $g_E$ is in the Borel subgroup   of $E_{11}$,     $g_l$ is formed from the generators of the $l_1$ representation
and  $g_h$ belongs to $I_c(E_{11})$. These group elements can be written in the form
$$
g_l= e^{z^A l_A}, \quad g_E=e^{A_{\bar \alpha} R^{\bar \alpha}}, 
\quad g_h= e^{\varphi \cdot S}
\eqno(1.2)$$
where $R^{\bar \alpha}$  and  $S^{\bar \alpha}$are the generators of the Borel subalgebra of $E_{11}$ and $I_c(E_{11})$ respectively. 
In   equation (1.2)  the  $z^A$ are the coordinates of the background space-time  and they depend on the parameters  $\xi^{\underline \alpha}$ of the brane world volume. The  fields 
$A_{\underline\alpha}$ are   the $E_{11}$ background fields, which include those of the maximal supergravity theories,    and they depend on the coordinates of the background spacetime
$z^A$. The fields $\varphi$ also depend on  $\xi^{\underline \alpha}$. The transformation $h$ of the local subgroup ${\cal H}$  depends on the parameters $\xi^{\underline \alpha}$ in an arbitrary way and we can use this to set some  of the  fields $\varphi$   to zero. We note that the non-linear realisation used for branes involves the additional fields $\varphi$  which are not present for the non-linear realisation used to derive the low energy effective action of strings and branes. 
\par
The Cartan forms are given by 
$$
{\cal V}= g^{-1} d g= {\cal V}_E+{\cal V}_l +{\cal V}_h^B, 
\eqno(1.3)$$
where 
$$
{\cal V}_E= g_E^{-1}dg_E , \ {\rm and }\ 
{\cal V}_l= g_E^{-1} g_h^{-1} (g_l^{-1}dg_l) g_h g_E  , \quad 
$$
$$
{\cal V}_h^B= g_E^{-1} (g_h^{-1}d g_h ) g_E= g_E^{-1}{\cal V}_h g_E= g_E^{-1} g_h^{-1} dg_h g_E(1.4)
\eqno(1.4)$$
The Cartan forms  ${\cal V}_E$  are just the Cartan forms of $E_{11}$ and they only depend on the background fields $A_{\underline \alpha}$.  The Cartan forms associated with the vector representation are given by   
$$
{\cal V}_l\equiv  \nabla^B  z^A l_A = g_E^{-1}g_h^{-1} (dz^Al_A ) g_h g_E = g_E^{-1} ( \nabla z^A  l_A )g_E\equiv    \nabla  z^\Pi  E_\Pi{}^Al_A
\eqno(1.5)$$
where $ \nabla\equiv d\xi ^{\underline \alpha } \nabla_{\underline \alpha }$ and 
$E_\Pi{}^A $ is defined by  $g_E^{-1} dz\cdot l g_E\equiv 
dz^\Pi E_\Pi{}^A l_A  $ which   is the vielbein in background spacetime and  also only depends on the  background fields $A_{\underline \alpha}$.   The fields $\varphi$ only  occur in the Cartan forms  $\nabla _{\underline \alpha} z^A$  and $g_h^{-1}d g_h$  which are  independent of the background fields $A_{\underline \alpha}$.
\par
The Cartan forms are inert under the rigid $g_0$ transformations, but under the local $h\in {\cal H}$ transformations they transform as   
$$ 
{\cal V}\to  h^{-1}{\cal V} h + h^{-1} d h
\eqno(1.6)$$
and in particular that 
$$
\nabla^B z^A l_A \to h^{-1} ( \nabla^B z^A l_A ) h ,  \quad 
{\cal V}_h^B\to  h^{-1}{\cal V}_h^B h + h^{-1} d h
\eqno(1.7)$$
Using this equation it is straightforward to explicitly compute the local transformations of the individual Cartan forms using  the $E_{11}\otimes _s l_1$ algebra. As the dynamics consists of  a set of equations that are invariant under these transformations and that the $\nabla_{\underline \alpha}^B z^A $ transform covariantly,  we are looking for equations which relate these Cartan forms to each other. We will also demand that the equations are invariant under arbitrary reparameterisations of the brane world volume. 
\par
For simplicity we will consider the  brane dynamics in the absence of background fields. In this case  we will take $g_E$ to be the identity element  and instead of taking the algebra $E_{11}\otimes _s l_1$  we take the non-linear realisation of $I_c (E_{11})\otimes _s l_1$ with local subalgebra ${\cal H}$. Then we only have the Cartan forms $ \nabla _{\underline \alpha} z^A $ and ${\cal V}_h = g_h^{-1} dg_h$. In fact the equations of motion are constructed from  only the Cartan forms $ \nabla _{\underline \alpha} z^A $. 
The brane dynamics  in the presence of the background fields can be readily found from the resulting equations by using equation (1.5) to   simply reinstate their presence by introducing the  vielbein in the way that this equation dictates, that is, make the replacement 
$\nabla _{\underline \alpha} z^A \ \to \ \nabla^B _{\underline \alpha }z^A $. We will not in this paper consider the construction of the Wess-Zumino term in the brane dynamics. 
 \par
 Although, there are still a number of features of the above  construction which are yet to be fully understood,  the general features are apparent [6,7]. 

\item{-} The  equations of motion  are constructed from objects that are first order in derivatives, that is, they involve the Cartan forms  $ \nabla _{\underline \alpha} z^A $ and not derivatives acting on these forms. As a result they are equations which equate the different  Cartan forms  associated with the vector representation to each other in such a way  as to preserve the local subalgebra ${\cal H}$. 

\item{-} Some of the equations which follow from the non-linear realisation are ones which  can be used to solve analytically for all of the fields $\varphi$ in terms of the coordinates, while  others  are dynamical equations for the coordinates and  these are duality conditions. 
\par
In this paper we will use the above guidelines to sketch the  dynamics of the low level branes in seven and eight dimensions.The advantage of this approach is that the reader can see what are the general features of branes in E theory without being distracted by the $E_{11}$ formalism and a morass of equations. We will, in particular, 
 focus on finding the generic form of the dynamical equations rather than the equations that are used to solve algebraically for the fields $\varphi$. 
 \par
  As just discussed  we are searching for duality relations between the Cartan forms which are  invariant under the transformations of the linear local subalgebra ${\cal H}$. We need not consider the rigid transformations as the Cartan forms are invariant under them. At the linearised level in the fields the equations of motion must be linear in the Cartan forms and as such we begin by searching for equations which relate one Cartan form to another. Indeed we must find relations that pair up the Cartan forms and relate them using the epsilon symbol on the world volume of the brane. For a p-brane we will divide the indices $\underline a, \underline b, \ldots = 0,1,\ldots , D-1$,  where $D$ is the dimension of the spacetime in which the brane moves, into those in the brane directions $a,b,\ldots =0,1,\ldots p$ and those transverse to the brane $a^\prime , b^\prime , \ldots =p+1,\ldots ,D-1$. As such the epsilon symbol in the brane world volume is denoted by $\epsilon^{a_1\ldots a_{p+1}}$. We will adopt this convention through out this paper. 
 \par
 We recall that the theory in $D$ dimensions arises when we delete the node labelled $D$ in the $E_{11}$ Dynkin diagram which leaves the algebra $GL(D)\otimes E_{11-D}$. We then  decompose $E_{11}$ into representations of this later algebra. These  representations are labelled by a level which depends on the node being deleted;  the level zero representation is 
 just the algebra that remains  when we delete the node labelled $D$ in the $E_{11}$ Dynkin, namely $GL(D)\otimes E_{11-D}$. 
 The Cartan involution invariant subalgebra of $E_{11}$, denoted $I_c(E_{11})$,  is of the form $R^{\alpha} - R^{-\alpha}$ where $\alpha$ is a positive root which has a positive level. Thus the generators in $I_c(E_{11})$  are made up of two generators of opposite level. The exception is  when these generators have level zero. We will refer to this as the level zero part of $I_c(E_{11})$   and it is the Cartan involution invariant subalgebra of $GL(D)\otimes E_{11-D}$ which contains the Lorentz algebra SO(1,D-1)   and 
 $I_c(E_{11-D})$ and is just $ SO(1,D-1)\otimes I_c(E_{11-D})$.  
 \par
The non-linear realisation used to construct the low energy effective action for string and branes requires a local subalgebra ${\cal H}$ which is a subalgebra of $I_c(E_{11})$. Indeed,  a p-brane breaks the SO(1,D-1) Lorentz symmetry of the background spacetime to $SO(1,p)\otimes SO(D-p-1)$ which is part of  ${\cal H}$.  A   brane charge also generically belongs to a representation of the U duality algebra,   $ E_{11-D}$,  and so also to $I_c(E_{11-D})$. As such the   particular brane charge which is active  will also break  $I_c(E_{11-D})$  into an algebra that is part of ${\cal H}$ and which we denote by ${\cal H}_0^U$. If we denote the level zero part of ${\cal H}$, in the above sense,  by ${\cal H}_0$, then ${\cal H}_0$ will contain $SO(1,p)\otimes SO(D-p-1)$ and ${\cal H}_0^U$;  indeed ${\cal H}_0= SO(1,p)\otimes SO(D-p-1)\otimes{\cal H}_0^U$.
 \par
A brane moves through a spacetime with the coordinates in the vector representation. In any dimensions the level zero coordinate  in the vector representation is $x^{\underline a}$ which  is the usual coordinate of our familiar spacetime. As we have just explained this must satisfy a duality relation with one of the other coordinates. If we consider a simple p-brane, that is, a brane whose charge in totally antisymmetric in its indices, then its charge will be of the form $Z^{a_1\ldots a_{p+1}}{}^{\bullet}$ where $\bullet$ denotes the indices of the representation of the U duality group $E_{11-D}$ to which it belongs. This representation breaks into representations of $I_c(E_{11-D})$. 
The particular brane charge that is active will break this latter group to  ${\cal H}_0^U$    under which the active charge is a singlet.   
Corresponding to this singlet charge the non-linear realisation possess a coordinate which we denote by $y_ {\underline a_1\ldots \underline a_{p+1}}$. We can write down a duality relation between this coordinate and the coordinate  $x^{\underline a}$ [7]
 $$
\nabla _\alpha x^a= -e_1 \epsilon^{a b_1 \ldots b_p} \nabla _\alpha y_{b_1\ldots b_p}
\eqno(1.8)$$
where $e_1$ is a constant. In all the cases studied so far one also finds that the non-linear realisation implies  the condition $\nabla_\alpha  x^{a^\prime}=0$.  Part of this equation  can be solved for a certain field $\varphi$ and the other part  is a dynamical equation for the transverse coordinates $x^{a^\prime}$ [7]. As a result one can derive  the equation 
$$
\sqrt {-\gamma} \gamma ^{\alpha \beta} \nabla _\beta x^{\underline a}= e_1 
\epsilon ^{ \alpha \beta \gamma_1\ldots \gamma _{p-1} }\nabla _\beta 
x^{\underline a \underline b_1\ldots \underline b_{p-1} } \nabla _{\gamma_1} x_{\underline b_{1}}\ldots \nabla _{\gamma_{p-1}} x_{\underline b_{p-1}}
\eqno(1.9)$$
where $\gamma _{\alpha\beta}= \nabla_\alpha x^a \nabla _\beta x^b \eta_{ab}$. This last equation   can be shown to be the familiar equations for brane dynamics and we refer the reader to reference [7] for an account. These equations hold for all the branes considered in this paper and so we will in what follows concentrate on the other dynamical equations. 
\par
The higher level symmetries of the non-linear realisation will transform the above equation (1.9)  into equations that are duality relations between the Cartan forms of   some the other coordinates in the vector representation. However, in this paper, rather than carry out such transformations we will simply search for generic duality relations that are  invariant under the local transformations of level zero, ${\cal H}^0$,  which consists of the groups   $SO(1,p)\otimes SO(D-p-1)$ and ${\cal H}_0^U$.  As it is a duality relation,  it will  contain the epsilon symbol $\epsilon^{a_1\ldots a_{p+1}}$. Let us consider  the  coordinates $x_{a_1\ldots a_n}$  which will have the Cartan form $\nabla _{\alpha} x_{a_2\ldots a_{n+1}}$. It will prove advantageous to rewrite this Cartan as $\nabla _{a_1} x_{a_2\ldots a_{n+1}}$ where 
$\nabla _{a}= (s^{-1})_a{}^\alpha \nabla_\alpha$ and $s_\alpha{}^a= \nabla_\alpha x^a$. We will also use this defintion through out this paper. 
We  expect that it will be  related to a coordinate with $m$ indices, that is  $x_{a_1\ldots a_m }$ where $m=p-n-1$ through an  equation whose generic form is given by 
$$
\nabla _{a_1} x_{a_2\ldots a_{n+1}}=e_2 \epsilon_{a_1a_2 \ldots a_{n+1}}{}^{b_{1}b_{2} \ldots b_{m+1}}\nabla_{b_{1}}x_{b_{2}\ldots b_{m+1}}
\eqno(1.10)$$
where $e_2$ is a constant. 
\par
The above discussion has neglected the fact that the coordinates $x_{a_1\ldots a_n}$  and $x_{a_1\ldots a_m }$ 
belong to the representations, denoted by $R^0_n $ and $R^0_m $ respectively  of the U duality group.  However, these representations decompose into a sum  of representations of the algebra  ${\cal H}_0^U$ preserved by the brane. To find a consistent equation  one must select representations in the two decompositions that can be related by an  ${\cal H}_0^U$ invariant tensor and use this tensor in the above duality relation. If $m=n$, that is when $2n= p-1$ the duality relation can become  a self duality relation. 
\par
The dynamics of the two and five brane in eleven dimensions, from the view point of E theory, was given in references [6,7]. It will be instructive to illustrate how  the above procedures apply to these branes. The brane charges in eleven dimensions are given in equation (1.2) and as a result  the non-linear realisation has the following coordinates 
$$
x^{\underline a}, x_{\underline a_1\underline a_2}, \ x_{\underline a_1\ldots \underline a_5}, \ x_{\underline a_1\ldots \underline a_7,\underline b},\  x_{\underline a_1\ldots \underline a_8},\ 
\  x_{ \underline b_1 \underline b_2  \underline b_3,  \underline a_1\ldots  \underline a_8},\ 
\  x_{( \underline c  \underline d ),  \underline a_1\ldots  \underline a_9},\ 
\  x_{ \underline c \underline d, \underline a_1\ldots  \underline a_9},\ 
$$
$$
\  x_{ \underline c, \underline a_1\ldots  \underline a_{10}}\ (2),\ 
x_{\underline a_1\ldots \underline a_{11}} ,\ 
x_{ \underline c,  \underline d_1\ldots  \underline d_4, \underline a_1\ldots  \underline a_9},\ 
x_{ \underline c_1\ldots  \underline c_6, \underline a_1\ldots  \underline a_8},\ 
x_{ \underline c_1\ldots  \underline c_5, \underline a_1\ldots  \underline a_9},\ 
x_{\underline  d_1, \underline c_1  \underline c_2  \underline c_3, \underline a_1\ldots  \underline a_{10}},\ (2),
$$
$$
\  x_{ \underline c_1 \ldots  \underline c_4, \underline a_1\ldots  \underline a_{10}},\ (2),\  x_{( \underline c_1 \underline c_2 \underline c_3 ),\underline a_1\ldots \underline a_{11}},\ x_{ \underline c, \underline b_1 \underline b_2, \underline a_1\ldots \underline a_{11}},\ (2),\  
x_{ \underline c_1\ldots  \underline c_{3},\underline a_1\ldots \underline a_{11}},\ (3),\ 
 \ldots  
\eqno(1.11)$$
The Cartan involution invariant subalgebra of $E_{11}$, denoted  $I_c(E_{11})$, has the Lorentz algebra SO(1,10) at level zero. 
\par
The charge of the M2  brane  is $Z^{a_1a_2}$  and selecting a particular charge  breaks the SO(1,10) Lorentz algebra  down to $SO(1,2)\otimes SO(8)$ which is the local subalgebra ${\cal H}$ at level zero, that is ${\cal H}_0=SO(1,2)\otimes SO(8)$. Corresponding to the M2 brane charge we have the coordinate  $x_{a_1a_2}$. This coordinate, together with the coordinate $x^a$ will obey the duality relation of equation of equation (1.9). This does indeed correctly describe the motion of the M2 brane. 
\par
The five brane has the brane charge $ \ Z^{a_1\ldots a_5}$ and this brane  breaks the Lorentz symmetry down to $SO(1,5)\otimes SO(5)$. The corresponding  coordinate $x_{a_1\ldots a_5}$, together with the coordinate $x^a$ obeys equation (1.9). At a lower level we have the two form coordinate $x_{a_1a_2}$ which, following the above discussion, should satisfy  a  self-duality relation which is invariant under  the level zero symmetries, that is, 
$$
\nabla_{[a_1}x_{a_2a_3 ]}={1\over 3!} \epsilon _{a_1a_2a_3}{}^{b_1b_2b_3} \nabla_{b_1}x_{b_2a_3}
  \eqno(1.12)$$
where 
$\nabla_a$ was defined above  equation (1.10). These equations we derived in reference [7] using the higher level symmetries in addition to those at level zero. They reproduce the correct dynamics to all orders in the usual embedding coordinate $x^a$ and up  to the  linear level in the world volume field $x_{a_1a_2}$; the situation for non-linear terms in the later field is discussed in reference [7]. These branes illustrate the general pattern,  the coordinates in the vector representation contain the usual embedding coordinate as well as the world volume fields and they  satisfy  duality relations. 
\medskip
{\bf 2 Branes in seven dimensions }
\medskip
The seven dimensional theory emerges when we decompose $E_{11}$ into 
$GL(7)\otimes SL(5)$ which is the algebra that emerges when we 
delete node seven in the $E_{11}$ Dynkin diagram. The generators in  the vector representation are given by [7]
$$ P_{\underline a}; \ \ Z^{MN} ; \ \ Z^{\underline a}{}_M; \ \ Z^{\underline a_1\underline a_2M}; \ \ Z^{\underline a_1\underline a_2\underline a_3}{}_{MN} ; \ \ Z^{\underline a_1\underline a_2\underline a_3,\underline b}, \ \ Z^{\underline a_1\ldots \underline a_4} , \ \ Z^{\underline a_1\ldots \underline a_4 M}{}_{N} , 
$$
$$
 \ \ Z^{\underline a_1\ldots \underline a_5 MN}, \ \ Z^{\underline a_1\ldots \underline a_5 (MN)} , \ \ Z^{\underline a_1\ldots \underline a_5}{}_{MN,P}, \ \ Z^{\underline a_1\ldots \underline a_4,\underline b MN}, \ldots   
\eqno(2.1)$$
where $\underline a , \underline b, \ldots =0,1,\ldots ,6$ and the indices $M,N,\ldots = 1,\ldots , 5$ are those of SL(5). 
\par
As a result the brane moves through a spacetime with the coordinates 
$$
x^{\underline a};\ \  x_{MN} ;\ \  
x_{\underline a}{}^M ;\ \   x_{\underline a_1\underline a_2M}  ;\ \  x_{\underline a_1\underline a_2\underline a_3}{}^{MN}  ;\ \   
x_{\underline a_1\underline a_2\underline a_3,\underline b} ;\ \  x_{\underline a_1\ldots \underline a_4}  ;\ \  
x_{\underline a_1\ldots \underline a_4 M}{}^{N} ;\ \  \ldots 
\eqno(2.2)$$
and the Cartan forms which belong to the vector representation  of the $E_{11}$ algebra  can be written in  the form 
$$
{\cal V}_l= \nabla x^{ \underline a} P_{\underline a} + \nabla x_{PQ} Z^{PQ}+
 \nabla x_{\underline a}{}^M Z^{\underline a}{}_M +  \nabla x_{\underline a_1\underline a_2M} Z^{\underline a_1\underline a_2M} + \nabla x_{\underline a_1\underline a_2\underline a_3}{}^{MN} Z^{\underline a_1\underline a_2\underline a_3}{}_{MN} 
$$
$$
+ \nabla x_{\underline a_1\underline a_2\underline a_3,b} Z^{\underline a_1\underline a_2\underline a_3,\underline b} +  \nabla x_{\underline a_1\ldots \underline a_4}  Z^{\underline a_1\ldots \underline a_4} + \nabla x_{\underline a_1\ldots \underline a_4 M}{}^{N}  Z^{\underline a_1\ldots \underline a_4 M}{}_{N} +\ldots 
\eqno(2.3 )$$
The Cartan forms transform under the local subalgebra ${\cal H}$ which is a subalgebra of $I_c(E_{11})$. At level zero  the latter   contains the Lorentz algebra SO(1,6) and the U duality algebra  $SO(5)$ and as such at level zero ${\cal H}$ is a subalgebra of this. 
We have already constructed the dynamics of the one and two branes in seven dimensions but it will be useful to illustrate  the methods of this paper to find their generic form. 
\medskip
{\bf 2.1 The one brane}
\medskip
The one brane has a two dimensional world volume and,  as explained above,  we take the indices in the directions of the world volume of the string to take the values  $a,b,\ldots =0,1$ and the reminder to be given by $a^\prime ,b^\prime ,\ldots =2,\ldots ,6$. The brane will preserve  $SO(1,1)\otimes SO(5)$ of the SO(1,6) Lorentz symmetry and so this symmetry belongs to  the local subaglebra ${\cal H}$. 
The one brane charge is given by $Z_M^{\underline a}$. A given  string has a given  charge and making this selection  we break the internal SO(5) symmetry down to SO(4) which will  also belong to ${\cal H}$. Indeed the level zero part is given by 
${\cal H}_0= SO(1,1)\otimes SO(5)\otimes SO(4)$.
As explained at the end of section one we adopt the duality relation (1.9)  between the  corresponding coordinate $y_{\underline a}$ and the usual spacetime coordinate $x^{\underline a}$. 
\par
Examining the coordinates of equation (2.1) we find the Lorentz scalar coordinates $x_{PQ}$  which under the decomposition to SO(4) leads to a coordinates in the $10= 6\oplus4$ representations of SO(4). We can denote these by $x_{ij}$ and $x_i$ ,where $ i,j=1,\ldots , 4$ respectively. Taking the former coordinate, we can  write down equations that are first order in the Cartan forms and are invariant under 
the local subalgebra ${\cal H}$ and in particular its level zero part, given above. Such equations are of the generic form 
$$
\nabla _a x_{ij} =  -
{1\over 2 }\epsilon_a{}^b \epsilon_{ijkl}  \nabla _b x^{kl}  \ \ \ { \rm or \ equivalently}\ \ \ \sqrt {-\gamma} \gamma ^{\alpha \beta}\nabla_\beta  x_{ij}= -
{1\over 2}\epsilon^{\alpha \beta}  \epsilon_{ijkl}  \nabla _\beta x^{kl} 
\eqno(2.1.1)$$
where $\nabla _a = (s^{-1})_a{}^\alpha \nabla_\alpha$ and $s_\alpha{}^a =\nabla_\alpha x^a$. Of course  to find the full equations of motion one has to find a set of equations that are invariant under the full symmetries of the non-linear realisation and not just the level zero symmetries. This was been done in reference [7] where   the choice of local subalgebra ${\cal H}$ and the corresponding transformations of the Cartan forms can be found. 
\par
Assuming that the equations that follow from the non-linear realisation imply that these are the only dynamical field we can count the number of bosonic degrees of freedom. We have $7-2=5$ degrees of freedom in $x^{a^\prime} $, taking into account the world volume reparameterisation symmetry,  and 
${4.3\over 2.2}= 3$ from $x_{ij}$ which gives us $8$ bosonic degrees of freedom. This is the number required for a half BPS brane that is maximally supersymmetric.  A brane with these degrees of freedom would arise from the dimensional reduction of the IIA string.
\medskip
{\bf 2.2 The two brane}
\medskip
The two brane has a three dimensional world volume and it breaks the Lorentz symmetry SO(1,6) into $ SO(1,2)\otimes SO(4) $ 
 which is in the local subalgebra.  The charge for the two brane is $Z^{a_1a_2 M}$ which transforms in the 5 of the internal SO(5) symmetry. Selecting a particular charge breaks SO(5) to SO(4). As a result the  local subalgebra at level zero is ${\cal H}_0= SO(1,2)\otimes SO(4)\otimes SO(4)$.  The active  two brane charge corresponds to a coordinate which we denote by $y^{\underline a_1\underline a_2}$ and this, together with the coordinate $x^{\underline a}$, 
will satisfy the duality relation of equation (1.9). 
\par
Examining the other coordinates in equation (2.2) we find the coordinates  $x_{MN}$ and  
$x_{\underline a}{}^M$ which, under the decomposition to SO(4),  decompose into a $10= 6 \oplus 4$ and a $5=4\oplus 1$ respectively. To find a duality relation  between the corresponding  Cartan forms we must choose the $4$ from each coordinate and then we can write down the relation 
$$
\nabla_a  x_i +{1\over 4} \epsilon _{a }{}^{c_1c_2} \nabla_{c_1} x_{c_2 i} =0 \ \ {\rm or \ equivalently} \ \ 
\sqrt {-\gamma} \gamma ^{\alpha \beta}\nabla_\beta  x_i+{1\over 4} \epsilon^{\alpha \beta_1\beta_2}\nabla_{\beta_1} x_d{}_i \nabla_ {\beta_2}x^d
=0 
\eqno(2.2.1)$$
\par
Assuming that these are the only active fields the number of bosonic degrees of freedom are 4 for $x^{a^\prime}$ and 4 for $x_i$ giving a total of 8 bosonic degrees of freedom. This is the correct number for a maximally supersymmetric brane in a type II theory. This is the same content as the dimensional reduced M2 brane of eleven dimensions. The fully non-linear equations of motion, the choice of local subalgebra ${\cal H}$ and the corresponding transformations of the Cartan forms were given in reference [7]. 
\medskip
{\bf 2.3 The three  brane}
\medskip
The three brane has a four dimensional world volume and   the part of the Lorentz symmetry SO(1,6) which is in the local subalgebra ${\cal H}$ is $ SO(1,3)\otimes SO(3) $. The charge for the three brane is $Z^{\underline a_1 \underline a_2 \underline a_3 MN}$ which belongs to the 10 dimensional representation of the internal symmetry SO(5). Choosing a particular charge for the three brane preserves only  $SO(3)\otimes SO(2)$ which belongs to the local subalgebra ${\cal H}$ at level zero. Decomposing the $10$ into this group we find that it consists of $10= (3,1)\oplus (3,2)\oplus (1,1)$, the last component being the active three brane charge. Let us denote the coordinate corresponding to this charge by $y_{\underline a_1 \underline a_2 \underline a_3}$ and this, together with the  coordinate $x^a$,  can be taken to obey the duality equation 
of equation (1.9). 
\par
Examining equation (2.2) we find the coordinate $x_a{}^{M}$ which belongs to the $5$ of SO(5). which can only be dual to itself given that the world volume epsilon symbol has  four spacetime indices . The $5$ decomposes into the $5=(3,1)\oplus (1,2)$ representations of $SO(3)\otimes SO(2)$. If we consider the latter we can write down the duality equation 
$$
\nabla_{[ a_1}x_{a_2] }{}^{i^\prime}= \pm {1\over 2} \epsilon_{a_1a_2}{}^{b_1b_2}\epsilon^{i^\prime j^\prime} \nabla _{b_1}x_{b_2 }{}_{j^\prime}
\eqno(2.3.1)$$
where $i^\prime, j^\prime, \ldots = 1,2$ are the SO(2) indices. There are no consistent duality relations one can write down for the $(3,1)$. 
\par
The coordinate $x_{MN}$ of equation (2.1) could be dual to the coordinate $x_{a_1a_2}{}_{M}$ which belong to the $10$ and $5$ of SO(5) respectively. Using the   decompositions of these representations given above we find that they have   $(3,1)$ in common which we denote by $x_{ij}$ and $x_{a_1 a_2}{}_i$ where $i,j\ldots = 1,2,3$ respectively. Using these coordinates we  can write down the duality relation  
$$
\nabla _{a_1}x_{ij}= \epsilon_{a_1}{}^{ b_1b_2b_3}\epsilon _{ijk} \nabla_{b_1}x_{b_2b_3} {}^k
\eqno(2.3.2)$$
\par
As with all the duality relations in this paper one can rewrite them so that they contain  $\nabla_\alpha$ rather than $\nabla_a$ using the techniques given at the end of section one. For equation (2.3.2) we find that it can be rewritten as 
$$
\sqrt {-\gamma} \gamma ^{\alpha \beta} \nabla _{\beta}x_{ij}= \epsilon^{\alpha \beta_1\beta_2\beta_3}\epsilon _{ijk} \nabla_{\beta_1}x_{d_1d_2} {}^k  \nabla _{\beta_2}x^{d_1} \nabla _{\beta_3}x^{d_2}
\eqno(2.3.3)$$
\par
Assuming that the other coordinates do not contribute to the dynamics then the number of bosonic degrees of freedom is 
$7-4=3$ for $x^{a^\prime}$, 3 for $x_{ij}$ and 2 for $x_a^k$ making 8 in all. Thus it contains  6 scalars and one vector as does $N=4$ supersymmetric Yang-Mills theory. We might consider this theory to be a dimensional reduction of the IIB D3 brane. 
\medskip
{\bf 2.4 The four  brane}
\medskip
The four brane has a five dimensional world volume and   the part of the Lorentz symmetry SO(1,6) which is in the local subalgebra ${\cal H}$ is $ SO(1,4)\otimes SO(2) $.  Examining equation (2.1) we find that there are three brane charges  with four indices only  two of which have four antisymmetric indices. We consider the case  that the four brane charge arrises from the charge $Z^{\underline a_1\ldots \underline a_4 M}{}_{N}$ which belongs to the $24=10\oplus 14$-dimensional representation of SO(5). We will choose the active brane charge to be $Z^{\underline a_1\ldots \underline a_4}{}^1{}_{ 2}$  which breaks SO(5) down to SO(3). Under which the $24$ contains the  four  singlets and one of these leads to a coordinate $y^{\underline a_1\ldots \underline a_4}$ which, together with $x^{\underline a}$, obeys equation (1.9).
\par
The Lorentz scalar coordinates are dual to the three form coordinates both of which belong to the 10 of SO(5) which decomposes into $10= 3\oplus 3\oplus 3\oplus 1$ of SO(3). We choose one of the 3 's and then write down the  generic duality  equation 
$$
\nabla_{a}x_i=\epsilon_{a}{}^{b_1\ldots b_4}\nabla_{b_1}x_{b_2b_3b_4} {}_i   \ \ \ i=1,2,3
\eqno(2.4.1)$$
The one form coordinates are dual to the two form coordinates and these both  belong to the 5 of SO(5). We choose one of the singlets under SO(3) and then we can  write down the generic equation 
$$
\nabla_{[ a_1}x_{a_2]}=\epsilon_{a_1a_2}{}^{b_1b_2}\nabla_{b_1}x_{b_2}
\eqno(2.4.2)$$
\par
We have $7-5=2$ degrees of freedom from the transverse coordinates, 3 from the Lorentz scalars and $5-2=3$ from the one form, which makes a count of 8 bosonic degrees of freedom. 
\par
For this case the duality  equations are not uniquely determined  by the level zero transformations of the local subalgebra and it is quite likely that there are other possible branes corresponding to the different choices of brane charge, the local subalgebra and the selection of different  representations of the internal symmetry.  What branes actually exist is  determined by the full symmetries of the local algebra. 
\medskip
{\bf 2.5 The five brane}
\medskip
The five brane has a six dimensional world volume and   the part of the Lorentz symmetry SO(1,6) which is in the local subalgebra ${\cal H}$ is $ SO(1,5)$.  Examining equation (2.1) we find that there are five charges with five Lorentz indices which could be the brane charge. We will consider that the active charge is $Z^{\underline a_1\ldots \underline a_5 (MN)} $ and in particular the component $Z^{\underline a_1\ldots \underline a_5 (11)} $ with corresponding coordinate $y_{a_1\ldots a_5}$. As a result the internal SO(5) symmetry is broken to SO(4) which belongs to the local subgroup. As a result ${\cal H}_0=   SO(1,5)\otimes SO(4)$. The coordinates $x^{\underline a}$  and 
$y_{\underline a_1\ldots \underline a_5}$ satisfies the duality relation of equation (1.9).  
\par
The Lorentz scalar coordinates $x_{MN}$ belong to the 10 of SO(5) which decomposes into the $10=6\oplus 4$ of SO(4). It is dual to one of the coordinates with four Lorentz indices. Two of these  are SO(5) singlets and the remaining one belongs to the  $24=10\oplus 14 $ of SO(5) which under SO(4) decomposes as $24= 6\oplus 9\oplus 4\oplus 4 \oplus1$. From these we take one of the $4$'s  and the $4$ from the Lorentz scalar coordinates and  then we can write down the generic duality relation
$$
\nabla_{a} x_i= \epsilon _{a}{}^{b_1\ldots b_5}\nabla _{b_1}x_{b_2\ldots b_5 i}\ ,\ \ i,j= 1,2,3,4
\eqno(2.5.1)$$
\par
The two form coordinate $x_{a_1a_2M}$ belongs to the $5$ of SO(5) which decomposes into $5=4\oplus 1$ under SO(4). Taking the singlet we find the self-duality relation which takes the generic form 
$$
\nabla_{[a_1}x_{a_2a_3]}=\pm {1\over 3!}\epsilon _{a_1a_2a_3}{}^{ b_1b_2b_3} \nabla_{b_1}x_{b_2b_3}
\eqno(2.5.2)$$
We have $7-6=1 $ transverse bosonic degrees of freedom, 4 from the scalars and ${4.3\over 2.2}=3$ from the two form making 8 in all. 
\par
For the case of the five brane the level zero transformations of the local subalgebra do not uniquely determine the duality relations. For example we could instead select the $6$-dimensional representation of SO(4) for the $x_{MN}$  and $x_{b_1\ldots b_4}{}^M{}_N$ and then write down a duality relation. This would contribute 6 bosonic degrees of freedom. We have also assumed that the $x_a{}^M$ and $x_{a_1a_2a_3}{}^{MN}$ do not satisfy a duality relation that leads to degrees of freedom. Indeed, if we had choose each to belong to the $4$-dimensional representation of SO(4) then we could have  written down a duality relation that contributes $4.4=16$ degrees of freedom. However, if one wants to get  only 8 bosonic degrees of freedom then one must adopt the possibilities we first gave. 
It is likely that there exist other   branes corresponding to different  choices and this will be resolved by see which putative brane  dynamics carries the full symmetries.


\medskip
{\bf 3 Branes in eight dimensions }
\medskip
We will now sketch the dynamics of some of the branes in eight dimensions; the pattern is similar to that in seven dimensions and so we will be brief. The eight dimensional theory emerges when we decompose $E_{11}$ into 
$GL(8)$ and the duality symmetry $SL(3)\otimes SL(2) $ which is the algebra that emerges when we 
delete node eight  in the $E_{11}$ Dynkin diagram. The generators in  the vector representation are given by 
$$ 
P_{\underline a} \ (1,1); \ \ Z^{i, j^\prime} \ (3,2) ; \ \ Z^{\underline a}{}_{i }  \ (\bar 3, 1); \ \ Z^{\underline a_1\underline a_2}{}^{i^\prime} 
 \ (1,2); \ \ Z^{\underline a_1\underline a_2\underline a_3}{}^ {i}  \ ( 3,1) ; 
, \ \ Z^{\underline a_1\ldots \underline a_4}{}_i{}^{j^\prime}   \ (\bar 3,2),
$$
$$
 Z^{\underline a_1\ldots \underline a_5}{}^{ i^\prime} {}_{j^\prime} \ (1,3) ,\ \ 
  Z^{\underline a_1\ldots \underline a_5}  \ (1,1), 
  Z^{\underline a_1\ldots \underline a_5}{}^{i }{}_{j}  \ (8,1), \ \ Z^{\underline a_1\ldots \underline a_6 i,j^\prime}  \ (3,2) , 
  \ \ Z^{\underline a_1\ldots \underline a_6}{}_{(ij),k^\prime} \ (\bar 6,2), \   \ldots   
\eqno(3.1)$$
where $\underline a , \underline b, \ldots =0,1,\ldots ,6$,  the numbers in the brackets refer to the representations of $SL(3)\otimes SL(2)$ that the generators belong to and $i,j,\ldots =1,2,3$ and $i^\prime,j^\prime,\ldots =1,2$ are the indices of SL(3) and SL(2) respectively. 
\par
As a result the brane moves through a spacetime with the coordinates 
$$
x^{\underline a} \ (1,1); \ \ x_{i, j^\prime} \ (\bar 3,2) ; \ \ x_{\underline a}{}^{i }  \ ( 3, 1); \ \ x_{\underline a_1\underline a_2}{}_{i^\prime} 
 \ (1,2); \ \ x_{\underline a_1\underline a_2\underline a_3}{}_ {i}  \ ( \bar 3,1) ; 
, \ \ x_{\underline a_1\ldots \underline a_4}{}^i{}_{j^\prime}   \ ( 3,2),
$$
$$
 x_{\underline a_1\ldots \underline a_5 }{}_{ i^\prime} {}^{j^\prime}  \ (1,3) ,\ \ 
 x_{\underline a_1\ldots \underline a_5}  \ (1,1), 
  x_{\underline a_1\ldots \underline a_5}{}_{i }{}^{j}  \ (8,1), \ \ x_{\underline a_1\ldots \underline a_6}{}_{ i , j^\prime}  \ (\bar 3,2) , 
  \ \ x_{\underline a_1\ldots \underline a_6}{}^{(ij),k^\prime} \ ( 6,2), \   \ldots   
\eqno(3.2)$$
The Cartan forms which belong to the vector representation  of the $E_{11}$ algebra  can be written in  the form 
$$
{\cal V}_l= \nabla x^{\underline a}  P_{\underline a} \ (1,1)+ \nabla   x_{i, j^\prime}    Z^{i, j^\prime} + \nabla  x_{\underline a}{}^{i } Z^{\underline a}{}_{i }  +\nabla   x_{\underline a_1\underline a_2}{}_{i^\prime}  Z^{\underline a_1\underline a_2}{}^{i^\prime} 
 + \nabla  x_{\underline a_1\underline a_2\underline a_3}{}_ {i}  Z^{\underline a_1\underline a_2\underline a_3}{}^ {i}  
 $$
 $$
 +\nabla  x_{\underline a_1\ldots \underline a_4}{}^i{}_{j^\prime} Z^{\underline a_1\ldots \underline a_4}{}_i{}^{j^\prime}  
 +\nabla   x_{\underline a_1\ldots \underline a_5 }{}_{ i^\prime} {}^{j^\prime}    Z^{\underline a_1\ldots \underline a_5}{}^{ i^\prime} {}_{j^\prime} +\nabla  x_{\underline a_1\ldots \underline a_5} 
  Z^{\underline a_1\ldots \underline a_5}  +\nabla 
  Z^{\underline a_1\ldots \underline a_5}{}^{i }{}_{j}  x_{\underline a_1\ldots \underline a_5}{}_{i}{}^{j} +\ldots 
\eqno(3.3 )$$
The Cartan forms transform under the local subalgebra ${\cal H}$ which is a sublagebra of $I_c(E_{11})$ which at level zero contains the Lorentz algebra SO(1,7) and the Cartan invariant subalgebra of the U duality algebra, that is,   $SO(3)\times SO(2)$.
\medskip
{\bf 3.1 The one brane}
\medskip
The one brane  will preserve  $SO(1,1)\otimes SO(6)$ of the SO(1,7) Lorentz symmetry.    The one brane charge is given by $Z_i^{\underline a}$ which belong to the $(3,1)$ representation of the  $SO(3)\times SO(2)$. Choosing a given   charge we break  the internal symmetry $SO(3)\times SO(2)$ down to $SO(2)\times SO(2)$ and we have the decomposition  $(3,1)=(2,1)\oplus (1,1)$. 
Denoting the coordinate associated with the later representation by  $y_{\underline a}$, we adopt the duality relation of equation (1.9) between this coordinate and the usual spacetime coordinate $x^{\underline a}$.
\par
 Examining the coordinates of equation (3.2) we find the Lorentz scalar coordinates $x_{i, j^\prime}$ in the  $ (\bar 3,2) $ representation of the internal symmetry $SO(3)\times SO(2)$; this decomposes into the  $(\bar 3,2) = (2,2) \oplus (1,2) $ representations of  $SO(2)\times SO(2)$. These  must satisfy a self-duality equation and we choose this to hold for the $(2,2)$ representation; 
$$
\nabla _a x_{i}{}_{j^\prime} =  -
{1\over 2}\epsilon_a{}^b \epsilon_{i}{}^{k}\epsilon_{j^\prime }{}^{ l^\prime}  \nabla _b x_{kl^\prime}  
\eqno(3.1.1)$$
where we define, as  before, $\nabla _a = (s^{-1})_a{}^\alpha \nabla_\alpha$ and $s_\alpha{}^a =\nabla_\alpha x^a$. In fact this is not the only equation we can write down which is invariant under the level zero symmetries. We can write the above equation but with no 
$ \epsilon$'s in the internal indices  and we could also write an equation for the $(1,2)$ representation, also with no $ \epsilon$'s in the internal indices 
Using the techniques discussed around equation (1.8) we can rewrite equation (3.1.1) in the form 
$$
\sqrt {-\gamma} \gamma ^{\alpha \beta}\nabla_\beta  x_{ij^\prime}= -
{1\over 2}\epsilon^{\alpha \beta} \epsilon_{i}{}^{k}\epsilon_{j^\prime }{}^{ l^\prime}   \nabla _\beta x_{kl^\prime} 
\eqno(3.1.2)$$ 
\par
Assuming that the equations that follow from the non-linear realisation imply that these are the only dynamical field we can count the number of bosonic degrees of freedom. We have $8-2=6$ degrees of freedom in $x^{a^\prime} $ and 
${2.2\over 2}= 2$ from $x_{ij^\prime}$ which gives us $8$ bosonic degrees of freedom. 
\medskip
{\bf 3.2 The two brane}
\medskip
The two brane preserves only  $ SO(1,2)\otimes SO(5) $ of the SO(1,7) Lorentz symmetry. The charge $Z^{a_1a_2 i^\prime}$ of  the two brane is  which transforms in the $(1,2)$ representation of the internal $SO(3)\times SO(2)$ symmetry and choosing a particular charge, say 
$Z^{a_1a_2 1}$,  breaks the internal symmetry down to  $SO(3)$. As a result ${\cal H}_0=  SO(1,2)\otimes SO(4)\otimes SO(3)$. We denote the coordinate associated with the active charge by  $y^{\underline a_1\underline a_2}$ and take this,  together with the coordinate $x^{\underline a}$,  to  satisfy the duality relation of equation (1.9). 
\par
Examining the other coordinates in equation (3.2) we find the Lorentz scalar coordinates  $x_{ii^\prime}$  in the $(3,2)$ representation of  $SO(3)\times SO(2)$. Under the decomposition to SO(3) it breaks into 
$(3,2)= 3\oplus 3$ while the dual coordinate $x_a^i$ belongs to the $(3,1)$ representation of $SO(3)\times SO(2)$ and it, of course, belongs to the 3 of SO(3). As a result we can write the duality equation 
$$
\nabla_{a} x_i=\epsilon_a{}^{b_1b_2} \nabla_{b_1}x_{b_2} {}_i \ , \ \ i=1,2,3
\eqno(3.2.1)$$
\par 
Assuming that these are the only active fields the number of bosonic degrees of freedom are $8-3=5$  for $x^{a^\prime}$ and 3 for $x_i$ giving a total of 8 bosonic degrees of freedom. 
\par
The two brane in eight dimensions was discussed in reference [13] and it would be interesting to see what is the relation to this work. 
\medskip
{\bf 3.3 The three  brane}
\medskip
The three brane has a four dimensional world volume and it breaks  the Lorentz symmetry SO(1,7) into  $ SO(1,3)\otimes SO(4) $. The charge for the three brane is $Z^{\underline a_1 \underline a_2 \underline a_3 }{}^i$ which belongs to the $( 3, 1)$ representation of 
the internal symmetry $SO(3)\times SO(2)$. Choosing a particular charge  preserves only  $SO(2)\otimes SO(2)$ and the $( 3, 1)$ representation becomes the representations $( 3, 1)= ( 2, 1)\oplus ( 1, 1)$. Denoting the coordinate associated with the last representation by $y_{\underline a_1 \underline a_2 \underline a_3}$, it,  together with the  coordinate $x^a$, obeys  the duality equation 
 (1.9). 
\par
Examining equation (3.2) we find the coordinate $x_a{}^{i}$. This  belongs to the $(3,1)$ of $SO(3)\otimes SO(2)$  which decomposes into 
$(3,1)= (2,1)\oplus (1,1)$. This field is self-dual and choosing the $(2,1)$ representation we can write down the equation 
$$
\nabla_{[a_1} x_{a_2 ]}{}_i= {1\over 2} \epsilon_{a_1a_2}{}^{b_1b_2}\epsilon_{i}{}^{j}\nabla _{b_1}x_{b_2}{}_j
\eqno(3.3.1)$$
There is no such consistent equation for the $(1,1)$ representation. 
\par
The coordinate $x_{ij^\prime}$ of equation (3.2) will  be dual to the coordinate $x_{a_1a_2}{}_{i}$ which belong to the $(3,2)$ and $(1,2)$ representations of $SO(3)\times SO(2)$ respectively. These    decomposes into the $(3,2)=(1,2)\oplus (2,2)$  and $(1,2)=(1,2)$ representations of $SO(2)\otimes SO(2)$. To write down a duality equation we must choose the in common $(1,2)$ representation and then 
$$
\nabla_{a} x_{i^\prime}= \epsilon_{a}{}^{b_1b_2b_3} \nabla_{b_1}x_{b_2b_3}{}_{i^\prime}
\eqno(3.3.2)$$
One could include an epsilon in the internal indices if required by the higher level symmetries. 
\par
Assuming that the other coordinates do not contribute to the dynamics then the number of bosonic degrees of freedom is 
$8-4=4$ for $x^{a^\prime}$, 2 for $x_{i^\prime}$ and $4-2=2$ for $x_{ai}$ making 8 in all.  
\medskip
{\bf 3.4 The four   brane}
\medskip
The four brane  breaks  the Lorentz symmetry SO(1,7) into  $ SO(1,5)\otimes SO(2) $. The charge for the three brane is $Z^{\underline a_1 \underline a_2 \underline a_3 \underline a_4}{}_i{}^{j^\prime}$ which belongs to the $( 3, 2)$ representation of 
the internal symmetry $SO(3)\times SO(2)$. Choosing a particular charge  preserves only  $SO(2)$ and the $( 3, 2)$ representation decomposes as  $( 3, 2)= 2\oplus 2 \oplus 1\oplus 1$. Let us denoting one of the singlet coordinates  by $y_{\underline a_1 \underline a_2 \underline a_3 \underline a_4}$. This coordinate together with the  coordinate $x^a$, obeys  the duality equation 
 (1.9). 
\par
The Lorentz scalar coordinates $x_{ij^\prime} $ belong to the $(3,2) $ representation of  $SO(3)\times SO(2)$ and which decomposes as given above. This coordinate is dual to the three form coordinate $x_{\underline a_1 \underline a_2 \underline a_3}{}_{i}$ which belongs to the $(3,1) $ representation of  $SO(3)\times SO(2)$. The two representations have in common the $2$ and $1$ representations 
If we choose the former representation we can write down the duality relation 
$$
\nabla_{a} x_i =\epsilon_{a}{}^{b_1b_2b_3b_4} \epsilon^{ij}\nabla _{b_1} x_{b_2b_3b_4}{}_j ,\ \ i,j=1,2
\eqno(3.4.1)$$
In fact one could  omit the epsilon in the internal indices and we could  also taken instead  the singlet at the symmetry level at which we are working. 
\par
The one form coordinate $x_{a}{}^i$ is dual to the two form coordinate  $x_{a_1a_2}{}_{i^\prime}$ which belong to the $(3,1)$ and $(1,2)$
representation of $SO(3)\times SO(2)$ respectively. When decomposed into $SO(2)$ they have in common only a singlet of SO(2) and we can write down the generic equation 
$$
\nabla_{[ a_1}x_{ a_2]}  =\epsilon_{a_1a_2}{}^{b_1b_2b_3}\nabla _{b_1} x_{b_2b_3}
\eqno(3.4.2)$$
\par
We have $8-5=3$ transverse bosonic degrees of freedom, $2$ in $x_i$ and $(5-2)=3 $ in $x_a$, making 8 in all.


\medskip
{\bf 4 The one brane in  four dimensions }
\medskip
We now briefly discuss the one brane in four dimensions. Decomposing $E_{11}$ into representations of $GL(4)\otimes E_7$ we find the  theory in four dimensions. The level zero part of $I_c(E_{11})$ is $SO(1,3)\otimes SU(8)$. The coordinates at low levels can be read off from  the table given earlier in this paper. It will be useful to present the coordinates in terms of representations of SU(8) 
$$
x^{\underline a} (1), x^{ij} (28) , x_{ij}  (\bar {28})  , x^a{}^{i}{}_{j} (63), x^a{}^{i_1\ldots i_4}  (70),  x^a ,\ldots  (1), \ \ i,j\ldots =1,\ldots , 8
\eqno(4.1)$$
where the number in the brackets gives the dimensions of the SU(8) representations. 
\par
The one brane preserves $SO(1,1)\otimes SO(2)$ of the SO(1,3) Lorentz symmetry. The brane charge belongs to the 
$63\oplus 70\oplus 1$ representations of SU(8). Let us choose the it to belong to the 70 dimensional representation, that is, 
$Z^   {a i_1\ldots i_4}$ and take the charge $Z^   {a 1234}$ to be the active charge. As a result the internal symmetry SU(8) gets broken to $SU(4)\otimes SU(4)$ with the decomposition  $70= (1,1)\oplus (1,1) \oplus (\bar 4 , 4)\oplus (6,6)\oplus (4,\bar 4) $ with  $Z^   {a 1234}$ being one of the singlets. We denote the corresponding coordinate by $y_{\underline a}$ and it, together with $x^{\underline a}$ will obey equation (1.9).
\par
The Lorentz scalar coordinates  decompose into representations of $SU(4)\otimes SU(4)$  as $28= (6, 1)\oplus (4,4)\oplus (1, 6)$ with similar results for the $\bar{28}$.  
These coordinates must  obey a self-duality condition, namely 
$$
\nabla_a x^{ij}= {1\over 2} \epsilon_{a}{}^{b} \epsilon^{ijkl} \nabla _b x_{kl}\ , i,j,k,l=1,2,3,4
\eqno(4.1)$$
\par
Counting the bosonic degrees of freedom we have $4-2=2$ from the transverse coordinates $x^{a^\prime}$ and ${4.3\over 2}=6$ from the 
$x^{ij}$ making 8 in all. There may well be  other possible one branes one could construct using the full symmetries of the non-linear realisation. 
\medskip
{\bf 5 Discussion }
\medskip
In two previous papers [6,7] we  discussed how to construct  brane dynamics as a non-linear realisation of $E_{11}\otimes _s l_1$ and we have shown that  it leads to many features of the brane dynamics that we know. While the construction of the dynamics of a given brane is rather complicated there have emerged a number of generic features; for example the dynamical equations when constructed from the Cartan forms  are a set of duality equations. By construction these equations   are invariant under the symmetries of the non-linear realisation and in particular the lowest level such symmetries.  In this paper we have applied these general features to find {\bf  in outline only} the dynamics of the low level  branes  in seven and eight dimensions. We find that the coordinates of the vector representation do indeed provided the fields for a set of duality relations that look to be of the right type in that they contain the expected coordinates describing how the brane moves through the usual spacetime as well as the required world volume fields. Unlike the superficial impression that might be gained by studying the  familiar branes in eleven and ten dimensions, world volume fields are generically present as long as they are  consistent with the duality relations  of equation (1.10). Hence although this paper contains no calculations of any length we hope that it does provide an insight into the general form of the dynamics of branes in E theory that is not obscured by the formalism or lengthy equations. 
\par
We find that the generic form of the dynamical equations are generally determined by the lowest level symmetries although for some of the higher level branes considered  here are several possibilities. It would be interesting to see how the generic dynamical equations become completely determined once they are required to be invariant under the higher level symmetries.
\par
While references [6,7] set out the general method to determine the brane dynamics from the non-linear realisation there are a number of steps where a very systematic path is absent. The situation is not unlike that which occurred when the  $E_{11}\otimes _s l_1$ non-linear realisation was used to find the low energy effective action for strings and branes, indeed it took quite a few  years before  the unique path became clear and the dynamics constructed. Three of the outstanding issues are as follows 

\item{-}The non-linear realisation requires for its construction a choice of local subalgebra ${\cal H}$ which for the branes is a subalgebra of the Cartan involution invariant subalgebra of $E_{11}$. Studies of particular branes have shown that the local subalgebras are much more subtle than one might naively expect [7]. It would be good to have a systematic way of choosing the local subalgebras that lead to brane dynamics; indeed such an understanding may lead to a classification of all branes from a purely algebraic viewpoint. We hope to report on  progress in this direction elsewhere. 

\item{-} The brane dynamics, as usually formulated,  consists of equations that contain second order derivatives acting on some of the fields. However, the brane dynamics that emerges from E theory is a set of duality equations that are first order in derivatives acting on fields. This is like the derivation of the low energy effective action of strings and branes from E theory. In the later case one can act on the duality equations with a spacetime derivative to eliminate field certain fields and obtain equations which are  second order in derivatives and are those of maximal supergravity once one restricts the equations  to the lowest level fields and coordinates. One would expect that a similar pattern will hold for the brane dynamics derived from E theory and it would be good to see that this is the case. In fact this is the case for branes with no world volume fields and at the linearised level in world volume fields   when they are present. However, it would be good to see the known features of the world volume fields emerge at the non-linear level. 
\item{}The equations that follow from the non-linear realisation contain  dynamical equations, as studied in this paper, and algebraic equations that solve for some of the $\varphi$ fields,  associated with the breaking of $I_c(E_{11})$ to ${\cal H}$,  in terms of derivatives acting on the coordinates. However, the dynamical equations also contain the $\varphi$ fields and so only once one has solved for these fields can on see the final form of the dynamical equation in terms of  the usual fields. A systematic way to do this has yet to be found. 
\item{-} The non-linear realisation for branes is formulated so that    the fields and coordinates depend on the brane parameters which label  points in the world volume swept out by the brane. However, in E theory the spacetime is infinite dimensional and so one can wonder what is the brane world volume in this very large spacetime? 
\par
The vector representations at low levels contains the brane charges of all branes that we are familiar with. However, it contains an infinite number of branes and it is reasonable to suppose that it encodes all brane charges. As a result E theory contains an infinite number of new degrees of freedom which might be very useful when studying problem such as black hole entropy. 
Once the above  issues are resolved it would be very interesting to determine the dynamics of the exotic branes which occur at higher levels in the vector representation.

\medskip
{\bf {Acknowledgements}}
\medskip
I wish to thank Michaella Pettit  and Paul Cook for discussions.   We wish to thank the SFTC for support from Consolidated grants number ST/J002798/1 and ST/P000258/1.

\medskip
{\bf {References}}
\medskip
\item{[1]} P. West, {\it $E_{11}$ and M Theory}, Class. Quant.  
Grav.  {\bf 18}, (2001) 4443, hep-th/ 0104081. 
\item{[2]} P. West, {\it $E_{11}$, SL(32) and Central Charges},
Phys. Lett. {\bf B 575} (2003) 333-342,  hep-th/0307098. 
\item{[3]} A. Tumanov and P. West, {\it E11 must be a symmetry of strings and branes },  arXiv:1512.01644. 
\item{[4]} A. Tumanov and P. West, {\it E11 in 11D}, Phys.Lett. B758 (2016) 278, arXiv:1601.03974. 
\item{[5]} P. West, {\it A Brief Review of E Theory},  The Proceedings of Abdus Salam's 90th Birthday meeting, 25-28 January 2016, NTU, Singapore, Editors L. Brink, M. Duff and K. Phua, World Scientific Publishing, IJMPA, {\bf Vol 31}, No 26 (2017)1750023,  arXiv:1609.06863
\item{[6]} P. West, {\it Brane dynamics, central charges and
$E_{11}$}, JHEP 0503 (2005) 077, hep-th/0412336.
\item{[7]} P. West, {\it E11, Brane Dynamics and Duality Symmetries },  Int.J.Mod.Phys. A33 (2018) no.13, 1850080, arXiv:1801.00669. 
\item{[8]} P. West, {\it Introduction to Strings and Branes}, Cambridge University Press, 2012. 
\item{[9]}  A. Kleinschmidt and P. West, {\it  Representations of G+++
and the role of space-time},  JHEP 0402 (2004) 033,  hep-th/0312247.
\item{[10]} P. West,  {\it $E_{11}$ origin of Brane charges and U-duality
multiplets}, JHEP 0408 (2004) 052, hep-th/0406150. 
\item{[11]} P. Cook and P. West, {\it Charge multiplets and masses
for E(11)}, ÊJHEP {\bf 11} (2008) 091, arXiv:0805.4451.
\item{[12]} O. Hohm and  H. Samtleben, {\it Exceptional Form of D=11 Supergravity}, Phys. Rev. Lett. 111 (2013)  231601, arXiv:1308.1673.
\item{[13]} V.  Bengtsson, M.  Cederwall, H.  Larsson and B. Nilsson.{\it U-duality covariant membranes } ,  JHEP0502 (2005) 020,  arXiv:hep-th/0406223. 

\end